\documentclass[manuscript,screen]{acmart}

\usepackage{xhfill}
\usepackage[skins]{tcolorbox}
\usepackage{varwidth}
\usepackage{graphicx,scalerel}

\usepackage{tabularx}

\makeatletter
\newtcolorbox{mybox}[1][]{
  enhanced,
  colframe=black, colback=white,
  sharp corners,
  boxrule=0.6pt,
  detach title,
  coltitle=black,
  colbacktitle=white,
  fonttitle=\bfseries, % Aggiungi \bfseries per il grassetto
  enlarge left by= 0mm,
  enlarge right by= 0mm,
  width=\linewidth,
  left=2mm,
  right=2mm,
  before upper=\setlength{\parindent}{17.62482pt}\everypar{{\setbox0\lastbox}\@minipagefalse\everypar{}},
  overlay={
    \node[anchor=east, 
          fill=tcbcolbacktitle, 
          font=\kvtcb@fonttitle]
          at ([xshift=-2mm]frame.south east) % Posiziona il nodo
    {
      \begin{varwidth}{\linewidth}
        \centering\tcbtitle\par
      \end{varwidth}
    };
  },#1}
\makeatother
\AtBeginDocument{%
  \providecommand\BibTeX{{%
    \normalfont B\kern-0.5em{\scshape i\kern-0.25em b}\kern-0.8em\TeX}}}

\setcopyright{rightsretained}
\copyrightyear{2024}
\acmYear{2024}
\acmDOI{XXXXXXX.XXXXXXX}

\acmConference[SE 2030]{International Workshop on Software Engineering in 2030}{July 15--16,
  2024}{Puerto Galinàs, Brazil}
\usepackage{xcolor, soul} 
\usepackage{paralist}
\newboolean{showcomments}
\setboolean{showcomments}{true} 
\ifthenelse{\boolean{showcomments}}
{
}

\begin{document}

\title{Engineering Digital Systems for Humanity: Challenges and Opportunities}% (Special Issue: 2030 Software Engineering Roadmap)}

\author{Martina De Sanctis}
\email{martina.desanctis@gssi.it}
\author{Paola Inverardi}
\email{paola.inverardi@gssi.it}
\author{Patrizio Pelliccione}
\email{patrizio.pelliccione@gssi.it}

\affiliation{
  \institution{\newline Gran Sasso Science Institute (GSSI), L'Aquila}
\streetaddress{Viale F. Crispi, 7}
  \city{LAquila}
  \state{Italy}
  \country{Italy}
  \postcode{67100}
}

\renewcommand{\shortauthors}{De Sanctis et al.}

\begin{abstract}
As testified by new regulations like the European AI act, the worries about the societal impact of (autonomous) software technologies are becoming of public concern. 
Social and human values, besides the traditional software behaviour and quality, are increasingly recognized as important for sustainability and long-term well-being.
In this paper, we identify the macro and technological challenges and opportunities of present and future digital
systems that should be engineered for humanity. Our specific perspective in identifying the challenges
is to focus on humans and on their role in their co-existence with digital systems. 
The first challenge considers humans in a proactive role when interacting with the digital systems, i.e., taking initiative in making things happening instead of reacting to events. The second concerns humans having an active role in the interaction with the digital systems i.e., on humans that interact with digital systems as a reaction to events. The third challenge focuses on humans that have a passive role i.e., they experience, enjoy or even suffer the decisions and/or actions of digital systems. Two further transversal challenges are considered: the duality of trust and trustworthiness and the compliance to legislation that both may seriously affect the deployment and use of digital systems.
\end{abstract}

\maketitle

\section{Introduction}

Digital systems are increasingly an integral part of our daily lives. They are pervasive and ubiquitous, automating our houses, the transportation means we use for travelling, and they support our work with various degrees of automation~\cite{InsightsISSRE}.
Consequently, there is a growing need for assurances of correct behavior to prevent undesired consequences.
The potential risks related to software are very clear in the case of critical systems, where the criticality can be related to safety, security, business (e.g., software controlling the stocks trading), or mission (e.g., software controlling satellites). However, nowadays risks are coming also from software not so obviously considered critical. This is the case of software automating documentation writing, personal assistants, news dissemination support, social network bots, human-resources support, investment assistance, etc. The development and popularity of AI, including ML, generative AI, and large language models, is triggering significant advancements in ways that were unthinkable only few years ago. 
The growing importance of technology is testified also by the increasing power that technological companies are having in spheres that go beyond their business. For instance, controlling communication in the earth or space brings enormous political power, which so far was reserved to countries and their governments. 

It is then becoming more and more clear that technology and software can have various and impactful consequences on our lives. Social and human values, besides the traditional software behaviour and quality, are increasingly recognized as important for sustainability and long-term well-being~\cite{pelliccione2023architecting}. This 
relates to  the amount of information that digital systems record about us, allowing for insights regarding every aspect of our lives ~\cite{cacmInverardi2019,InsightsISSRE} that go beyond privacy as far as 
predictions of our future choices and decisions are concerned. 

In this paper, we identify the macro and technological challenges and opportunities of present and future digital systems that should be engineered for humanity. The importance of the topic is testified also by the establishment of regulations, starting from {\em GDPR} in 2018 to the recent {\em AI act}, the first law in the world to regulate the use of AI, which has been recently approved and released by the European Parliament~\cite{AI_Act_briefing}.
Our specific perspective in identifying the challenges is therefore to focus on humans and on their role and/or position in their co-existence with digital systems.  

The first challenge 
focuses on humans adopting a proactive role when interacting with the digital systems, i.e., taking initiative in making things happening instead of reacting to events that happen. This relates to humans being able to {\em continuously program systems}, i.e., to program systems at design time but also being able to alter, reprogram, or change the system behaviour during runtime. This calls for languages and ways to program systems (e.g., not necessarily by writing code, but also by examples, through feedback, or through voice) that are accessible to everyone despite of its background or knowledge.
The second challenge concerns humans having an active role in the interaction with the digital systems. Indeed, there can be an overlap with the continuous programming of the previous challenge. However, in this challenge we focus on the ``reactive'' aspect of the human interaction, i.e., on humans that interact with digital systems as a reaction to events.
The third challenge focuses on humans that have a passive role but that experience, enjoy or suffer in the worst cases the decisions and/or actions of digital systems. For instance, this is the case of AI-based system supporting human resources (HR) offices, banking systems in the decision for granting a loan or insurances, and so on.
The fourth and last challenge concerns the duality of trustworthiness and trust. Trustworthiness concerns the designing of system in order to behave safely and guarantee security or quality aspects. Instead, trust concerns the acceptability of systems from the point of view of humans.

\section{State of the Art}
\label{sec:motivation}

The various industrial revolutions have brought about transformations in all societal systems. The evolutionary history of industry has now reached Industry 5.0. 
Its main objective 
is to prioritize human well-being within manufacturing systems, to foster prosperity for the sustainable development of all humanity~\cite{Industry5.0_EU_whitePaper}.
Moving towards this new perspective is challenging, and the manufacturing paradigm needs time to adapt to the requirements of a novel society~\cite{LENG2022279}. 
Society~5.0 has also been proposed~\cite{HUANG2022424} aiming at 
balancing economic advancement with the resolution of social problems. 

Responsible computing in the digital world has been also thoroughly examined~\cite{cacmInverardi2019}. The author delved into the foundational principles of addressing ethical concerns in autonomous systems to mitigate the harm of digital society. 
Along the same line, Lu et al. identified a roadmap~\cite{DBLP:conf/cain/00010X0X22} as well as best practices~\cite{lu2023responsible} for responsible AI systems. Responsible AI concerns with ensuring the responsible development and operation of AI systems. It has emerged as a significant challenge in the AI era targeting both legal and ethical aspects that must be 
considered to achieve trustworthy AI systems. 
In ~\cite{DBLP:conf/cain/00010X0X22}, the authors 
defined a roadmap about software engineering
methods on how to develop responsible AI systems, including requirements engineering, systems design and operation. 
The roadmap also refers to
ensuring the compliance of the development
and use of AI systems to ethical regulations
and responsibilities, and defining architectural styles for responsible-AI-by-design.

The need for architecting and engineering
value-based ecosystems was also highlighted in~\cite{pelliccione2023architecting}. In this work, the authors develop the concept of (dynamic) ecosystems as emerging from the mutual interaction of people, systems, and machines, due to the continuous digitalization of the human world. These interactions can have different nature, such as collaborative, competitive, or malicious, and they can be further enhanced by AI. In this context, humans could find themselves vulnerable in their engagements with the digital realm, due to issues concerning societal values, such as fairness and privacy. To ensure a better digital society, the authors argue the need for engineering values-based by design digital ecosystems, 
equipped with built-in mechanisms to protect social and human values.

Empowering the user with personalized software connectors that are able to reflect user's moral values in the interactions with digital systems is the approach taken in the Exosoul project \cite{Autili2019}. The approach has been applied to privacy profiles \cite{DBLP:journals/ase/RuscioIMN24,DBLP:journals/csr/InverardiMP23} and to more general ethical profiles \cite{DBLP:conf/hhai/AlfieriIMP22}.
In line with this work, Boltz etal.~\cite{BoltzSEAMS2024}, in their vision paper, emphasize the need of human empowerment within the framework of self-adaptive socio-technical systems, which demand mechanisms for balancing diverse needs, values, and ethics at the individual, community, and societal levels. 
Other approaches offer a different perspective for operationalising values. Bennaceur etal.~\cite{icse_BennaceurHNZ23} promote the  shifting of the definition and measurement of users values from the early stages of software development to the runtime, supported by the software itself.
Under the broad umbrella of dynamic ecosystems, we can categorize research about the human–machine teaming paradigm~\cite{HenryKSLGWWMS22,ClelandHuangCZCAV24}. 
While systems are still expected to operate autonomously, they are viewed as partners rather than tools in accomplishing mission objectives. To enable this transition, humans and machines must engage in closer interaction, fostering meaningful partnerships where decisions are collaboratively~made~\cite{ClelandHuangCZCAV24}.

Multiple countries worldwide are formulating and enacting legislation and policies concerning AI governance. In the USA Biden enacted the “Executive Order on the Safe, Secure, and Trustworthy Development and Use of Artificial Intelligence''~\cite{BidenOrder}, whose purpose is that of promoting the responsible use of AI. 
Similarly, the Cyberspace Administration of China issued the final version of the “Interim Administrative Measures for Generative Artificial Intelligence Service''~\cite{ChinaOrder}, establishing measures applying to the provision of generative AI services  
within China.
The European Union has been the precursor to these legislations. Indeed, the European Parliament released the so-called {\em AI Act}, the first law to regulate the use of AI in Europe~\cite{AI_Act_briefing}. 
It provides a standardized framework for both the use and provision of AI systems within the European Union (EU). This framework delineates different requirements and obligations tailored on a risk-based approach and used to classify AI systems. The AI act applies primarily to providers and deployers of AI systems and general purpose AI models. 
Four categories of risk are identified and legal interventions are tailored to these concrete level of risk. 
AI systems presenting  
threats to people's safety, livelihoods and rights, are prohibited. 
Analogously to what happened with privacy and the European GDPR regulation, the EU has pioneered with the AI Act, and yet multiple countries worldwide  
are designing AI governance legislation and policies. 
As a result of the phenomenon known as the \emph{Brussels effect}~\cite{BrusselsEffect}, entities end up complying with EU laws even beyond its borders.

Overall, the analyzed works delineate the current operational landscape of modern digital systems and shed light on the challenges they entail, as discussed in Section~\ref{sec:challenges}.

\section{Motivating Examples}
\label{subsec:examples}
Before describing the macro and technological challenges and opportunities, we introduce motivating examples that will be exploited later as running examples.

%\noindent{\textbf
\subsection{User-friendly Cobots}%}
This first example refers to the use of Cobots, i.e, collaborative robots intended for direct human-robot interaction within a shared space, or where humans and robots are in close proximity, in a production environment. Our cobot is equipped with software enabling the programming of the robot, i.e., the specification of mission to be accomplished. 
The cobot provides multiple instruments to facilitate its programming by domain experts, who may not necessarily be experts in robotics or ICT. The first instrument is a textual and graphical domain specific language (DSL) to enable a more ``traditional'' way of programming the robot. The second instrument allows programming the cobot using voice commands. Through the third instrument the cobot learns by examples, either with the human showing by movements the desired behaviour or by enabling learning from feedback on incorrect actions. The programming is not ended before the mission execution but it is a continuous activity while the cobot and the human are working together. 
In addition to mechanisms ensuring the safe and secure behaviour of the robot, such as a safety-monitored stop and power and force limitation when necessary, the cobot is ethically aware to ensure the fulfilment of ethical standards during its operations.

%\noindent \textbf
\subsection{Assistive robots}
This motivating example is inspired by~\cite{askarpour2021robomax}. 
In nursing homes,  
assistive robots should first establish short conversations based on the user's conditions to figure out the overall user's health status, and afterwards, they bring pills and a glass of water. Robots should record this activity to enable a caregiver to assess whether the patient has taken the pills during the 
\emph{interaction} with a robot. 
Robots should occasionally perform regular check-ups on people with particular conditions  
and, if they require any assistance, alert the nurse.
Robots might reduce social isolation, and the risk of distress and confusion in older people~\cite{sharkey2014robots}. They do not show distress, anger, and the worst sides of human behaviour, but they are also incapable of real compassion and empathy or understanding~\cite{sharkey2014robots}.
Consequently, robots 
should be engineered so as to avoid the risk of reducing rather than improving the quality of life for older people. This has to do with privacy concerns, including the right to be let alone, the right to secrecy, and control of personal data~\cite{muller2020ethics}, as well as with human dignity. Specifically,  
robots should monitor the behaviour of older people and recognize reluctant complaints and distresses that may be caused by the interaction with robots and the absence of humans. In this case, robots should be capable of delegating control to humans,  
to accommodate needs, such as human compassion or empathy.

%\noindent{\textbf
\subsection{AI for banking systems}%} 
Digitalization and AI are increasingly used in financial services, such as banking and insurance, as also reported by
Moody's,  
in its recent global study on attitudes, adoption, and use cases for AI in the realm of risk management and compliance~\cite{moodys_report}.  
Today customers can easily login into their mobile banking app, check their account balance and perform many other tasks. At the same time, AI algorithms used by the financial service provider analyze their transactions, spending habits, and financial goals.
These technologies are used for disparate reasons, e.g., improving efficiency, increasing speed, and saving costs~\cite{moodys_report}.
Based on the collected  data, the AI system can personalize the user experience by offering tailored financial advice, such as budgeting tips or investment opportunities. 
From one side, AI enhances the banking experience by providing personalized recommendations and robust security measures to ensure customer satisfaction. However, from the other side, these technologies are exposed to fairness issues that might be due to, e.g., bias in  data and/or AI models. For instance, loans granting can be biased by the customers nationality, gender, etc., without even being aware of the presence and influence of such biases. Then, these systems should be engineered so as to avoid unfair behaviour, privacy violations, and biases~\cite{GuidottiMRTGP19,MehrabiMSLG21}.

\section{Macro and technological challenges and opportunities}
\label{sec:challenges}

\begin{figure}[!htb]
     \centering
\includegraphics[width=0.6\textwidth]{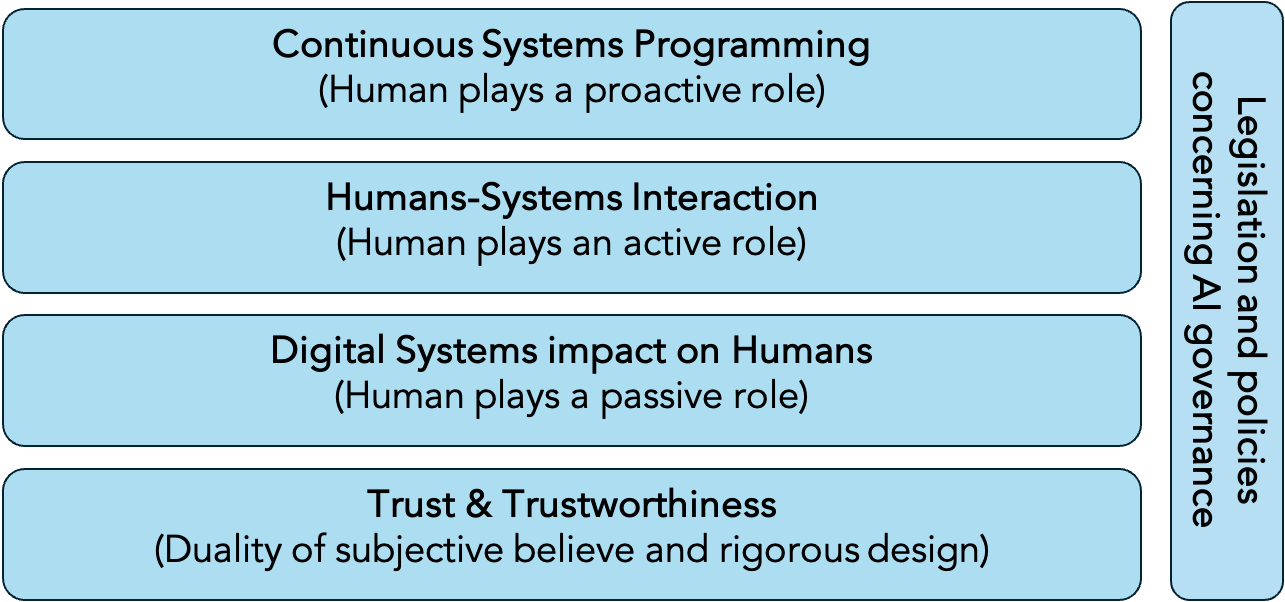}
     \caption{Main Challenges for Engineering Digital Systems for Humanity}
     \label{fig:challenges}
     \vspace{-.5cm}
\end{figure}

Based on the context set up 
and considering the examples discussed in Section~\ref{subsec:examples}, we envisage 
four macro and technological challenges, as shown in  
Figure \ref{fig:challenges}. The figure shows also legislation and policies concerning AI governance that will play a crucial role in the near future. Connected to each challenge, we also highlight the opportunities.

\subsection{Continuous Systems Programming} 
\label{subsec:programming}

In digital systems 
humans can play a \textbf{proactive role}, i.e.,  
humans can create or control a situation, e.g.,~drive/change the systems' behaviour, rather than just responding to it. 
First, humans should be able to program system, i.e.,~to specify what they should do in the specific situations in which they are used. It is important to highlight that 
this programming phase occurs independently from the system's programming phase during production. It is a post-production programming phase carried out in the context of use.
It can go from the settings of some parameters to a richer specification of the system behaviour, what should be avoided, and the specification of quality and ethical aspects~\cite{Autili2019,Dennis_Bentzen_Lindner_Fisher_2021,Svegliato_Nashed_Zilberstein_2021}. More often than not, the persons that should accomplish this programming phase are not developers and, therefore, the programming language should be accessible and easy to use by stakeholders that are experts of the domain but not experts in ICT. 
As an example, we can consider the user-friendly cobots or assistive robots motivating examples. In the case of the cobots, the humans that should program the cobots are experts of the domain in a production environment but they are not necessarily experts in robotics. The language should be user-friendly and intuitive but at the same time unambiguous and rigorous, and should use the same terminology of the domain. In the case of assistive healthcare robots, the language should enable the patient, or operators supporting the patient in this activity, to specify preferences like consensus about the treatment, risk management assessment, balance for benefit or moral attitude.

In the literature we can find approaches trying to make the specification of missions accessible to users not experts in ICT, while enabling them to correctly and accurately specify the mission robots should accomplish. Patterns for mission specification have been identified by surveying the state of the art and by formulating catalogs of specification patterns for mobile robots~\cite{Menghi2021,Menghi2022}.  
Catalogs come with a structured English grammar enabling the specification of the mission in English. This user-friendly specification of the mission is then automatically transformed into a temporal logic specification, thus enabling the use of various tools for, e.g., synthesizing planners or controllers or to enable verification. Other researchers have proposed domain specific languages (DSLs) with the aim of proposing user-friendly ways for programming robots~\cite{Dragule2021,Swaib2021}. 
Other approaches envision gesture-based robot programming in the context of human-robot collaboration in shared workplaces~\cite{NUZZI2021102085}, and cobot programming for collaborative industrial tasks~\cite{ELZAATARI2019162}. In~\cite{NUZZI2021102085}, the authors present a ROS-based software called MEGURU where
a user-friendly gesture language is developed using gestures that are easy for users to perform and convey a clear, intuitive command.
A survey on intuitive cobot programming is instead reported in~\cite{ELZAATARI2019162}.

\begin{mybox}[colframe=black, title=CH1: Continuous Systems Programming]

\noindent\textbf{Challenge:}
Engineering for humanity requires a proper balance between simplicity/flexibility/accessibility and rigorousness/unambiguity/accuracy to enable humans, non-expert in ICT, to confidently and correctly program systems in a continuous fashion.   
Regulations concerning AI governance, such as the AI Act, envisage also a post-market monitoring of the use and provision of AI systems, on a risk-based basis. When continuous systems programming comes into play, a continuous compliance with such regulations and policies must be guaranteed, both to protect humans playing with a proactive role, but also to prevent malicious behaviours from humans. 

\noindent\textbf{Opportunities:}  
there is the need of languages specific of the domain that should be simple and intuitive but also rigorous to guide the user towards the correct specification
of their needs. Other means to program systems should be also explored, e.g., imitation learning to enable the programming through examples, or also exploiting large language models to program systems via voice and speech.
A successful and effective solution would require the involvement of a multi-disciplinary team able to cover competencies that go beyond the pure technical aspects. 
\end{mybox}

\subsection{Humans-Systems Interaction}
\label{subsec:interaction}

In digital systems 
humans can play an \textbf{active role}, i.e., 
humans are actively engaged with the systems, and interact with them, but they react to events instead of being proactive. 
In scenarios equivalent to that of Assistive Robots, humans and systems share the same environment, not necessarily physical, but also only digital or hybrid. 
This implies that besides privacy concerns, other ethical values and human dignity come into play. {\em Digital ethics} is the branch of ethics concerning the study and evaluation of moral problems related to data and information, algorithms, and corresponding practices and infrastructures, with the aim of formulating and supporting morally good solutions~\cite{floridi2018soft}. Digital ethics is made by 
{\em hard ethics} and {\em soft ethics}, which are sometimes intertwined inextricably. System producers may already adhere to the hard ethics rules defined by legislation, which are considered collectively accepted values. 
Soft ethics define personal preferences, 
and it is the responsibility of the soft ethics to shape the users' interaction with the digital world~\cite{cacmInverardi2019,Autili2019}. The subjectivity in the ethics specification is also confirmed by the results of the moral machine experiments, which highlights how demographic and cultural traits affect the moral preferences~\cite{MoralMachine}.
These considerations highlight the need for  
engineering 
solutions for flexible, customizable, and privacy preserving interactions between humans and systems. 
Humans should be able to express and negotiate their preferences, while systems should dynamically adhere to the diverse soft ethics of the users they interact with. This calls for emphatic systems, i.e., systems able to understand, interpret, and respond to human emotions~\cite{stark2021ethics}. 
In other words, there is the need of instruments to guarantee that the behavior of the system, e.g., robot(s), will be compliant with the specified ethical preferences (as discussed in Section~\ref{subsec:programming}).
The use of synthesis techniques can then help to automatically generate the correct-by-construction logic needed for coordinating the robots and their interactions with humans, as well as the environment, in a way that the specified mission is accomplished in the correct and morally good manner~\cite{cacmInverardi2019,Autili2019}. Synthesis techniques can take as input the mission specification, including the ethical preferences, and can automatically generate a controller able to mediate the interactions between robots and humans so to guarantee the accomplishment of the mission, when possible, while protecting and preserving ethical preferences. Also, the controller should be able to understand the situations in which a redistribution of control from robots to humans is needed to not violate human ethics~\cite{EUAIExpertGroup}. Adjustable autonomy is the means to redistribute the operational control
among different parts of the system, as well as humans~\cite{mostafa2019adjustable}. 
Alfieri etal.~\cite{Alfieri_HHAI24} investigated the literature with the aim of clarifying the distinction between human replacement and human augmentation, in the context of autonomous intelligent systems.  
They observed a prevailing negative perception regarding human replacement, whereas there is a generally favorable attitude towards enhancing them.

\begin{mybox}[colframe=black, title=CH2: Humans-Systems Interaction]

\noindent\textbf{Challenge:} Elicitation of ethics is an activity that cannot be completely anticipated at design time because it is hard to accurately profile humans according to their moral preferences. AI is a promising instrument to infer emotions and ethical preferences in a continuous fashion. However, this instrument could be itself too risky for human beings. In fact, the recent AI act of the European community {\em ``[...] prohibits placing on the market, putting into service [...], or use of AI systems to infer emotions of a natural person in the areas of workplace and education institutions except in cases where the use of the AI system is intended for medical or safety reasons."}~\cite{AI_Act_briefing}.   
Also, the development of systems able to automatically redistribute autonomy and pass the control is challenging.

\noindent\textbf{Opportunities:} there are various research opportunities connected to these challenge, from the continuous elicitation and specification of ethics, to the engineering of systems able to measure and redistribute autonomy and the control, to instruments able to exploit the specification for, e.g., synthesizing ethical-aware mediators to let the system’s hard and soft ethics match the user’s soft ethics.

\end{mybox}

\subsection{Digital Systems impact on Humans}
\label{subsec:impact}

In digital systems 
humans can play a merely \textbf{passive role}, in the sense that the interaction with systems does not involve 
active participation. These types of digital systems require
to be engineered in order to guarantee fair and ethical decision-making processes and comply with ethical standards and regulations.
This is a crucial aspect since their users might be vulnerable and unaware that their values could be at risk, like in the AI for banking system example. One aspect that requires attention is removing bias from systems (e.g., processing, design, algorithm itself) and from the data used to train AI-based systems~\cite{ChakrabortyMM21,MehrabiMSLG21,https://doi.org/10.48550/arxiv.2207.07068}. This is important because bias could lead to unfair behaviours of the systems with potential discrimination for individuals, groups or subgroups of people~\cite{Das33}. 
Enhancing explainability of AI and systems' behaviour is widely investigated to provide greater transparency~\cite{GuidottiMRTGP19,dwivedi2023explainable,TantithamthavornCHC23,Martinez-Fernandez22}. However, in the literature, we can find various terms that often overlap each other, such as explainability, transparency, interpretability, and understandability. It would be beneficial to clarify and have a clear explanation for each of these concepts~\cite{10.1007/978-3-030-82017-6_8}, similarly to what is done for more traditional quality attributes~\cite{ISOstandard}.
Transparency is also highlighted in the AI Act~\cite{AI_Act_briefing}, which provides various facets of transparency: (i) AI systems should be developed and used in a way that allows appropriate traceability and explainability, (ii) humans should be made aware that they communicate or interact with an AI system, (iii) deployers should be made aware by providers of the capabilities and limitations of the AI system, and (iv) affected persons should be made aware about their rights (such as the rights to free movement, non-discrimination, protection of private life and personal data).

\begin{mybox}[colframe=black, title=CH3: Digital Systems impact on Humans]

\noindent\textbf{Challenge:} this challenge covers many aspects, including explainability, transparency, interpretability, and understandability. More clarity of concepts in order to avoid misunderstandings would be beneficial. The AI Act is also introducing new requirements for high-risk and limited-risk AI-based systems which go sometimes over the more technical aspects, as for instance the transparency requirement of informing users about their rights and the fact that they are using an AI system. In the near future, there will be the need of comprehensive solutions to engineer digital systems for humanity.

\noindent\textbf{Opportunities:} there are various research opportunities connected to these challenge, from bias identification, mitigation, and removal, explainability, transparency, etc. The community could benefit from a new quality standard that includes also these emerging new qualities of AI-systems that are becoming unavoidable.

\end{mybox}

\subsection{Trust \& Trustworthiness}
\label{subsec:trust}

Trust and Trustworthiness are a cross-cutting challenge that are already explored in the previous challenges. In fact, the importance of \emph{trust} and \emph{trustworthiness} within these systems is increasingly compelling, as these
are central aspects, especially with respect to humans interplay with digital systems.
Here we would like to highlight the dichotomy between the terms trust and trustworthiness, which are frequently treated as synonyms, despite their distinct meanings. 
As discussed in~\cite{lu2023responsible},
trust concerns with the subjective acceptability of systems from the humans' perspective, while 
trustworthiness concerns the designing of system in order to behave safely and guarantee security or quality aspects.

\noindent\textbf{Trust.} Explainability, transparency, interpretability, and understandability~\cite{10.1007/978-3-030-82017-6_8} are qualities of the system that promote trust.   
Also, trust can be affected by their replacing or augmenting the role of autonomy and humans~\cite{Alfieri_HHAI24}.
The before mentioned adjustable autonomy as a means to redistribute the operational control among systems and humans~\cite{mostafa2019adjustable}, is also motivated by the need for ways to deal with the absence of trust and confidence of human users.

\noindent\textbf{Trustworthiness.} 
Autonomous systems are considered trustworthy when their design, engineering and operation mitigates potentially harmful outcomes~\cite{NaisehBR22}. This can depend on many factors such as accountability, robustness, adaptation ability in dynamic and uncertain environments, security against attacks, and so on. Common to the various known techniques for trustworthiness demonstration, from synthesis to verification and testing, there is the need to formulate specifications for validating them~\cite{AbeywickramaBCDKLMMMNRRWWE24}. 
However, specifying modern AI-powered systems for the purpose of demonstrating their trustworthiness poses new challenges. For instance, testing AI systems is challenging due to their complex nature that makes conventional testing approaches insufficient or impractical for these systems~\cite{NeelofarA24}.

\begin{mybox}[colframe=black, title=CH4: Trust \& Trustworthiness]

\noindent\textbf{Challenge:} the considerations above highlight the need for engineering and operationalize digital systems in such a way to increase the human confidence in these systems and facilitate the human acceptance. This calls for ways to measure the level of trust, and concepts such as replacement and augmentation, to foster trust. Humans should be able to clearly understand the systems capabilities and responsibilities, while systems should behave transparently.
Additionally, there is the need for adequate testing approaches for AI-powered systems relying on equally adequate specifications of such systems, able to capture their complex nature in all its facets.

\noindent\textbf{Opportunities:} among the research opportunities connected to this challenge, we mention the need to identify the gap between processes, technologies and skills used for realising traditional systems and those required for engineering trust and trustworthy systems. Also, opportunities cover how these systems should be specified for the purpose of being trusted over time and supporting trustworthiness validation and verification.

\end{mybox}

\section{Conclusions}
\label{subsec:roadmap}

Digital systems are increasingly pervasive and part of our lives. 
Human beings are at risk, due to, e.g., 
lack of transparency, unfair behaviour, and biases of digital systems. 
To protect humans in their interplay with these systems, 
we believe that new engineering approaches are needed for the design and operation of digital systems for humanity. For this purpose, in this paper we focused on 
the perspective of humans and their role in their co-existence with digital systems. This perspective adds further criticalities into known challenges, such as human-system interaction or trustworthiness, previously approached solely from a systems standpoint. This work aims to shed light on an additional dimension that both researchers and practitioners should consider while advancing research on engineering modern digital systems. By presenting challenges alongside opportunities, we provide insights into areas deserving further exploration.

\section*{Acknowledgements}
This work has been partially funded by 
\begin{inparaenum}[(a)]
\item the MUR (Italy) Department of Excellence 2023 - 2027, 
\item the PRIN project P2022RSW5W - RoboChor: Robot Choreography, and 
\item the PRIN project 2022JKA4SL - HALO: etHical-aware AdjustabLe autOnomous systems.
\end{inparaenum}
%

%\newpage
\bibliographystyle{ACM-Reference-Format}
\bibliography{biblio}

%\appendix

\end{document}